\documentclass[unnumsec,webpdf,contemporary,large]{oup-authoring-template}%

\graphicspath{{Fig/}}

\usepackage{amsmath,amssymb,amsfonts}
\usepackage{longtable}
\usepackage{hyperref}
\usepackage{xurl}
\usepackage{tabularray}
\usepackage{lscape}
\usepackage{booktabs} 
\theoremstyle{thmstyleone}%
\usepackage{mdframed}
\theoremstyle{thmstyletwo}%
\theoremstyle{thmstylethree}%

\begin{document}

\journaltitle{Bioinformatics}
\DOI{}
\copyrightyear{2025}
\pubyear{2025}
\access{}
\appnotes{Application Note}

\firstpage{1}

\title{PhenotypeToGeneDownloaderR: automated multi-source retrieval and validation of phenotype-associated genes}

\author[1,2,$\ast$]{Muhammad Muneeb}
\author[1,2,$\ast$]{David B. Ascher}

\authormark{Muneeb et al.}

\address[1]{\orgdiv{School of Chemistry and Molecular Biology}, \orgname{The University of Queensland}, \orgaddress{\street{Queen Street}, \postcode{4067}, \state{Queensland}, \country{Australia}}}
\address[2]{\orgdiv{Computational Biology and Clinical Informatics}, \orgname{Baker Heart and Diabetes Institute}, \orgaddress{\street{Commercial Road}, \postcode{3004}, \state{Victoria}, \country{Australia}}}

\corresp[$\ast$]{Corresponding authors: David B. Ascher, Email: \href{email:d.ascher@uq.edu.au}{d.ascher@uq.edu.au}; Muhammad Muneeb, Email: \href{email:m.muneeb@uq.edu.au}{m.muneeb@uq.edu.au}}

\abstract{
\textbf{Motivation:} Identifying phenotype-associated genes is a common first step in polygenic risk score construction, enrichment testing, target prioritisation and variant interpretation, but relevant evidence is distributed across heterogeneous databases with different interfaces, formats and evidence models. \textbf{Results:} We present PhenotypeToGeneDownloaderR, a phenotype-guided R/Python pipeline for automated gene retrieval, harmonisation, symbol validation and cross-source summary analysis. Given a phenotype term, the pipeline queries integrated biological databases, standardises per-source outputs, combines gene lists, validates retrieved symbols against the NCBI human gene reference and generates summary tables and visualisations. Across 13 clinically relevant phenotypes and 13 databases, PhenotypeToGeneDownloaderR generated 136,487 raw gene retrievals, with at least one source returning genes for every phenotype. Across all 13 phenotypes, 100,175 of 114,345 combined input symbols were retained after direct or synonym-based validation, corresponding to an 87.6\% validation rate. Cross-source overlap was low, supporting the complementarity of integrated evidence sources. Against an HPO/ClinVar/OMIM-derived gold standard, the pipeline recovered 1,039 of 1,056 known phenotype-associated genes, corresponding to 98.4\% recall. PhenotypeToGeneDownloaderR provides a lightweight, reproducible upstream framework for generating candidate gene sets for downstream prioritisation and interpretation. \textbf{Availability and implementation:} PhenotypeToGeneDownloaderR is implemented in R and Python, released under the MIT licence, and available at \url{https://github.com/MuhammadMuneeb007/PhenotypeToGeneDownloaderR}. \textbf{Supplementary information:} Supplementary data are available online.
}

\keywords{phenotype-associated genes, phenotype-guided retrieval, multi-source integration, gene symbol validation, candidate gene prioritisation, bioinformatics software}

\maketitle
 % =========================
% INTRODUCTION
% =========================

\section{Introduction}
Identifying genes associated with a phenotype or disease is a frequent starting point for human genetics and bioinformatics analyses, including enrichment testing, polygenic risk score construction, target prioritisation and variant interpretation \cite{Uffelmann2021,Mills2019,Witte2010}. However, phenotype--gene evidence is distributed across resources that differ in scope, interface, terminology, evidence type and output format. Clinical resources such as ClinVar \cite{Landrum2020}, OMIM \cite{Amberger2019} and HPO \cite{Kohler2021} capture curated disease and phenotype associations; pathway and protein resources such as KEGG \cite{Kanehisa2023}, Reactome \cite{Gillespie2022}, STRING \cite{Szklarczyk2023} and UniProt \cite{UniProt2023} capture functional or interaction evidence; and integrative resources such as Open Targets \cite{Ochoa2023}, DisGeNET \cite{Pinero2020} and the GWAS Catalog \cite{Sollis2023} provide disease, genetic association and multi-evidence links. As a result, researchers often need to query multiple resources independently and manually reconcile heterogeneous gene identifiers before downstream analysis.

Existing packages and web resources are valuable, but many focus on a single database, require pre-defined gene sets, or emphasise downstream enrichment rather than phenotype-first retrieval and harmonisation \cite{Magno2019gwasrapidd,Cao2023,Murphy2021MungeSumstats}. This creates a practical bottleneck: the same phenotype may yield different candidate gene sets depending on which source is queried, how terms are matched and how gene symbols are reconciled.

PhenotypeToGeneDownloaderR addresses this gap by providing a unified phenotype-first workflow for multi-source gene retrieval, standardisation and cross-source evidence aggregation. Given a phenotype term, the pipeline retrieves candidate genes from integrated biological databases, writes standardised per-source outputs, combines gene lists, validates symbols against the NCBI human gene reference and produces reproducible summary outputs for downstream prioritisation. The tool is intended as an upstream triage layer: it generates analysis-ready candidate gene sets and source-support summaries, rather than claiming to establish causal gene--phenotype relationships.

\section{Materials and Methods}

PhenotypeToGeneDownloaderR consists of two linked components: an R-based retrieval layer and a Python-based downstream analysis layer. Given a phenotype term, the retrieval layer queries integrated biological databases and writes standardised per-source CSV outputs. The downstream layer combines cross-source results, summarises recovery patterns, and generates analysis-ready outputs. Retrieved gene symbols are additionally cross-referenced against the NCBI human gene reference to distinguish valid symbols from artefacts and to resolve synonyms to current official symbols.

\begin{figure}[!ht]
  \centering
  \includegraphics[width=\columnwidth]{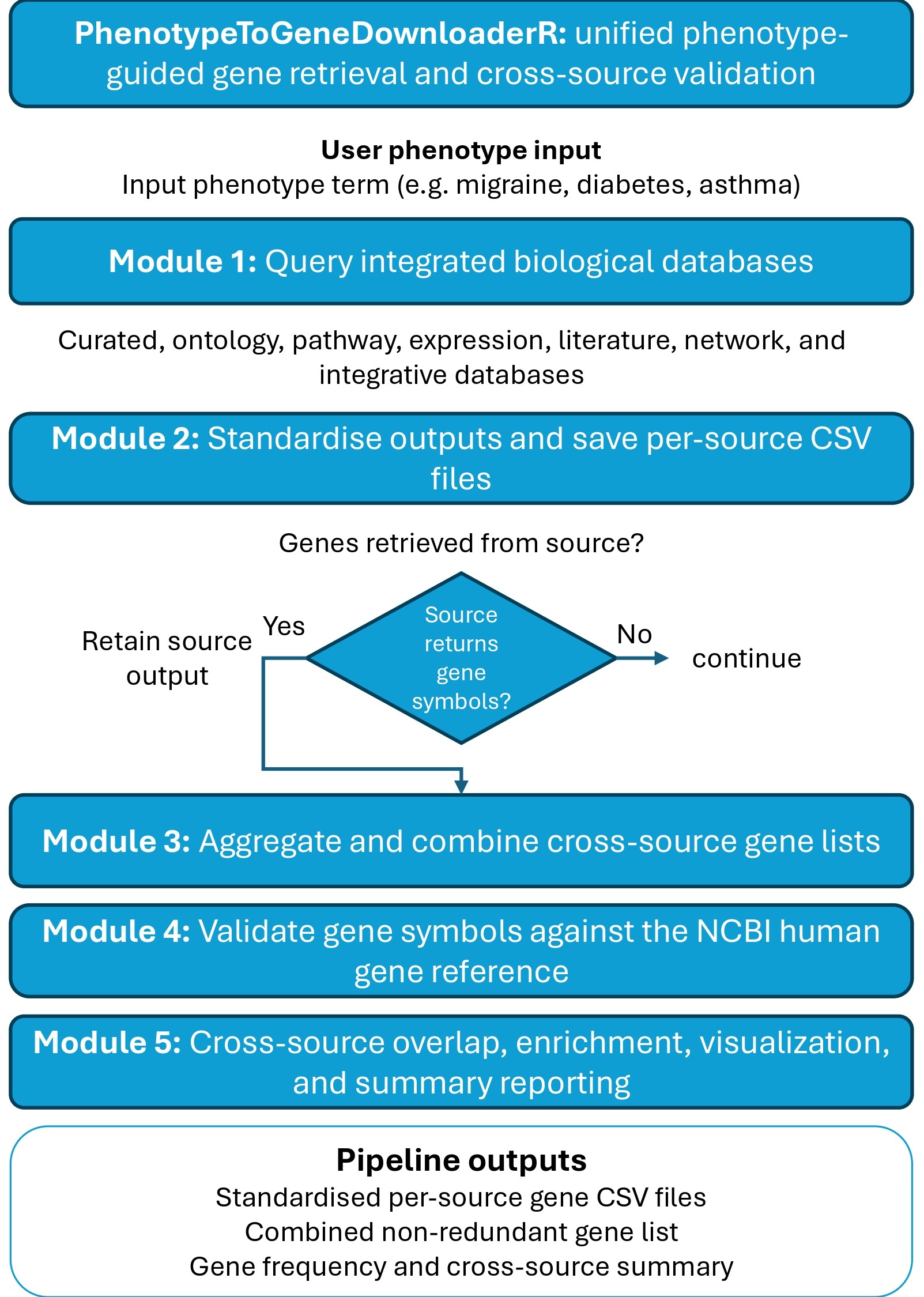}
  \caption{Overview of the PhenotypeToGeneDownloaderR workflow. A phenotype term is used to query integrated biological databases, generate standardised per-source CSV outputs, combine cross-source gene lists, validate gene symbols against the NCBI human gene reference, and produce downstream summary analyses and visualisations.}
  \label{flowchart}
\end{figure}

For the present benchmark, the pipeline was evaluated across 13 clinically relevant phenotypes and 13 integrated biological databases spanning literature, clinical genetics, ontology, pathway, expression, protein-function, and integrative evidence types. The overall workflow is shown in Figure~\ref{flowchart}.

% =========================
% RESULTS AND DISCUSSION
% =========================
 
\section{Results and Discussion}

Across 13 clinically relevant phenotypes and 13 integrated biological databases, PhenotypeToGeneDownloaderR generated 136,487 raw gene retrievals, with at least one source returning genes for every phenotype. Twelve of 13 databases returned genes for at least one phenotype, while OMIM and GTEx achieved complete phenotype coverage in this benchmark. Open Targets and the GWAS Catalog contributed the largest total gene yields, whereas lower-yield resources such as Reactome, KEGG and Gene Ontology contributed more selectively, reflecting differences in database scope, terminology, query behaviour and evidence model (Table~\ref{tab:source_summary_main}; Supplementary Section~S1).

Gene-symbol validation and harmonisation retained 100,175 of 114,345 combined input symbols across all 13 phenotypes, corresponding to an overall validation rate of 87.6\%. This step removed non-gene artefacts and unresolved identifiers while rescuing outdated or alternative aliases through synonym-based mapping to current official symbols. In total, 4,912 validated symbols were resolved through synonym mapping, representing 4.9\% of the validated set (Supplementary Section~S2). These results show that the pipeline can convert heterogeneous source-level outputs into cleaner candidate gene sets suitable for downstream review and interpretation.

KEGG and DisGeNET require cautious interpretation in the current benchmark. KEGG returned source-level records for some phenotypes, but these produced zero validated genes after the current parsing and symbol-validation step, indicating an implementation-level parsing limitation. DisGeNET returned no genes across the evaluated phenotypes because API access was required but was not available during these runs. These results are therefore reported as access or implementation limitations rather than evidence that these resources lack relevant phenotype--gene associations.

Cross-source overlap was low overall, supporting the complementarity of the integrated evidence sources. Most genes were supported by a single database, while a smaller subset was recovered by multiple independent sources and may represent higher-confidence candidates for downstream prioritisation. GWAS Catalog, GTEx, Open Targets and OMIM contributed substantial unique gene fractions, indicating that each source captures partly distinct biological or evidence dimensions (Supplementary Section~S3).

To assess recovery of established phenotype-associated genes, we compared the retrieved gene sets against an HPO/ClinVar/OMIM-derived gold standard. PhenotypeToGeneDownloaderR recovered 1,039 of 1,056 curated known phenotype-associated genes, corresponding to 98.4\% recall. Precision@20 analysis further supported the ability of the source-frequency ranking to prioritise known phenotype-associated genes among the highest-ranked candidates, with a mean Precision@20 of 64.1\% across phenotypes with non-empty gold-standard sets (Supplementary Section~S4). Downstream enrichment using g:Profiler produced significant enrichment results for all 13 phenotypes, supporting the biological coherence of the generated candidate gene sets while remaining a downstream characterisation rather than causal validation (Supplementary Section~S5).

The benchmark also showed practical usability. Total runtime across 13 phenotypes was 94.1 minutes, with a mean runtime of 7.2 minutes per phenotype and modest peak memory use. Empirical comparison with implemented modules showed that the full pipeline recovered a larger fraction of gold-standard genes than the GWAS Catalog-only or Open Targets-only modules, while the top 500 source-ranked genes submitted to g:Profiler retained 793 of 1,056 gold-standard genes (75.1\%) (Supplementary Section~S6). Together, these results support the use of PhenotypeToGeneDownloaderR as a lightweight and reproducible first-pass framework for phenotype-guided candidate gene retrieval and cross-source evidence aggregation. The tool is not designed to prove causal gene--phenotype relationships directly; rather, it provides harmonised, source-supported candidate gene sets for downstream interpretation, enrichment analysis and expert review.

\begin{table}[!ht]
\centering
\caption{Compact summary of per-database retrieval performance across 13 phenotypes. Success indicates the number of phenotypes for which a database returned at least one gene symbol. Total genes indicates the sum of unique genes returned across all phenotypes prior to cross-source merging and symbol validation.}
\label{tab:source_summary_main}
\small
\begin{tabular}{lccr}
\toprule
Database & Success & Rate (\%) & Total genes \\
\midrule
Open Targets  & 11/13 & 84.6  & 47489 \\
GWAS Catalog  & 8/13  & 61.5  & 41750 \\
OMIM          & 13/13 & 100.0 & 22918 \\
PubMed        & 12/13 & 92.3  & 10038 \\
GTEx          & 13/13 & 100.0 & 7149 \\
ClinVar       & 11/13 & 84.6  & 2849 \\
HPO           & 7/13  & 53.8  & 1975 \\
UniProt       & 10/13 & 76.9  & 1090 \\
KEGG          & 5/13  & 38.5  & 417 \\
Reactome      & 5/13  & 38.5  & 407 \\
STRING-DB     & 6/13  & 46.2  & 262 \\
Gene Ontology & 2/13  & 15.4  & 143 \\
DisGeNET      & 0/13  & 0.0   & 0 \\
\bottomrule
\end{tabular}
\end{table}

\section{Availability and implementation}

PhenotypeToGeneDownloaderR is freely available as an open-source software pipeline at \url{https://github.com/MuhammadMuneeb007/PhenotypeToGeneDownloaderR}. The repository is released under the MIT licence and includes the complete R-based phenotype-to-gene retrieval scripts, Python-based downstream analysis scripts, dependency files, documentation and example usage instructions. The current release described in this manuscript is version 1.0.0. The retrieval layer is implemented in R and coordinated through \texttt{download\_genes.R}, which executes source-specific database modules for phenotype-guided gene retrieval. The downstream analysis layer is implemented in Python and includes scripts for source coverage analysis, gene-symbol validation, cross-source overlap assessment, known-gene recovery, enrichment analysis and summary visualisation. Installation requirements are provided through \texttt{requirements.R}, \texttt{requirements.txt} and \texttt{environment.yml}, enabling users to reproduce the R and Python environments required for execution. Given a phenotype term, the pipeline generates per-source output files, combined cross-source gene lists, source-summary tables, validation outputs, overlap statistics and publication-ready summary plots. Example commands, expected output structure and documentation are provided in the GitHub repository. The benchmark phenotypes and supplementary outputs described in this manuscript provide test cases for reproducing retrieval coverage, symbol validation, cross-source complementarity, known-gene recovery and runtime analyses. PhenotypeToGeneDownloaderR is intended as a reproducible upstream candidate-gene retrieval and prioritisation framework for downstream analyses such as enrichment testing, polygenic risk score development, target prioritisation and variant interpretation. It does not infer causal gene--phenotype relationships directly; instead, it harmonises and summarises evidence from multiple public biological databases to support downstream review and interpretation.

\section{Competing interests}

The authors declare that they have no competing interests.

\section{Author contributions statement}

M.M. wrote the first draft of the manuscript and wrote, tested, and documented the code. M.M. analysed the results. D.A. reviewed and edited the manuscript. All authors contributed to the methodology.

\section{Data availability}

The source code, documentation, example commands and example output files are available in the GitHub repository at \url{https://github.com/MuhammadMuneeb007/PhenotypeToGeneDownloaderR}. The underlying databases are publicly accessible: PubMed (\url{https://pubmed.ncbi.nlm.nih.gov/}), OMIM (\url{https://www.omim.org/}), ClinVar (\url{https://www.ncbi.nlm.nih.gov/clinvar/}), HPO (\url{https://hpo.jax.org/}), Gene Ontology (\url{https://geneontology.org/}), KEGG (\url{https://www.kegg.jp/}), Reactome (\url{https://reactome.org/}), STRING (\url{https://string-db.org/}), GTEx (\url{https://gtexportal.org/}), UniProt (\url{https://www.uniprot.org/}), Open Targets (\url{https://www.opentargets.org/}), DisGeNET (\url{https://www.disgenet.org/}), and the GWAS Catalog (\url{https://www.ebi.ac.uk/gwas/}).

\section{Acknowledgments}

Not applicable.

\bibliographystyle{unsrt}
\bibliography{reference}

\end{document}

\renewcommand{\thetable}{S\arabic{table}}
\renewcommand{\thefigure}{S\arabic{figure}}
\maketitle

\section*{Supplementary Section S1: Pipeline execution and source coverage}

The following tables report gene retrieval and execution statistics for the PhenotypeToGeneDownloaderR pipeline applied to 13 clinically relevant phenotypes across 13 integrated biological databases. Supplementary Table~S1 provides per-phenotype per-database gene counts. Supplementary Table~S2 summarises per-database retrieval performance aggregated across all phenotypes. Supplementary Table~S3 reports combined all-source validation results per phenotype. Supplementary Table~S4 provides benchmark runtime and memory usage. Supplementary Table~S5 presents overall pipeline statistics.

\subsection*{Supplementary Table S1: Unique genes retrieved per phenotype and database}

A value of zero indicates that the database returned no genes for that phenotype, that the source-specific output file was not available, or that no recognised gene-symbol column was detected in the source output. Counts reflect unique gene symbols after within-source deduplication and before cross-source merging or symbol validation. OT = Open Targets; GWAS = GWAS Catalog; STRING = STRING-DB; GO = Gene Ontology.

\begin{landscape}
\begin{table}[!ht]
\centering
\caption{Unique gene symbols retrieved per phenotype and database. Counts reflect unique gene symbols after within-source deduplication but before cross-source merging or symbol validation.}
\label{tab:s1_count_matrix}
\scriptsize
\resizebox{\linewidth}{!}{%
\begin{tabular}{|l|r|r|r|r|r|r|r|r|r|r|r|r|r|}
\hline
\textbf{Phenotype} &
\textbf{ClinVar} &
\textbf{GTEx} &
\textbf{GWAS} &
\textbf{HPO} &
\textbf{KEGG} &
\textbf{OMIM} &
\textbf{OT} &
\textbf{PubMed} &
\textbf{Reactome} &
\textbf{STRING} &
\textbf{UniProt} &
\textbf{DisGeNET} &
\textbf{GO} \\
\hline
Asthma                          & 288 & 616 &  5597 & 181 &  32 & 2185 & 7443 & 1244 &   0 & 50 & 105 & 0 &   0 \\
Blood pressure medication       &   0 & 200 &     0 &   0 &   0 &  901 &  253 &   37 &  22 &  0 &   0 & 0 &   0 \\
Body mass index                 & 328 & 528 & 19286 &   0 &   0 & 1342 & 6185 &  947 & 116 &  0 &  48 & 0 &   0 \\
Cholesterol lowering medication &   0 & 322 &     0 &   0 &  52 & 1364 &    0 &   54 &  99 &  0 &   0 & 0 &   0 \\
Depression                      & 556 & 559 &     0 & 475 &  60 & 2538 & 6829 & 1041 &   0 & 50 & 183 & 0 & 112 \\
Gastro-oesophageal reflux       &  15 & 574 &     0 &   0 &   0 & 2843 & 1626 &  270 &   0 &  0 &   0 & 0 &   0 \\
Allergic rhinitis               & 114 & 612 &  1428 &  13 &   0 & 2294 &    0 &    0 &   0 & 50 &  16 & 0 &   0 \\
High cholesterol                &  96 & 724 &     0 &   0 &  52 & 1019 & 2431 & 1143 & 135 & 12 &   6 & 0 &  31 \\
Hypertension                    & 427 & 573 &  4279 & 650 &   0 & 1424 & 9766 & 1291 &   0 & 50 & 496 & 0 &   0 \\
Hypothyroidism                  & 275 & 691 &  5026 & 428 &   0 & 2019 & 3050 & 1048 &   0 &  0 & 103 & 0 &   0 \\
Irritable bowel syndrome        &  10 & 538 &  1168 &   0 & 221 & 1952 & 1959 &  780 &  35 &  0 &   2 & 0 &   0 \\
Migraine                        & 382 & 613 &  2595 & 161 &   0 &  789 & 2523 &  828 &   0 &  0 &  45 & 0 &   0 \\
Osteoarthritis                  & 358 & 599 &  2371 &  67 &   0 & 2248 & 5424 & 1355 &   0 & 50 &  86 & 0 &   0 \\
\hline
\textbf{Total}                  & 2849 & 7149 & 41750 & 1975 & 417 & 22918 & 47489 & 10038 & 407 & 262 & 1090 & 0 & 143 \\
\hline
\end{tabular}%
}
\end{table}
\end{landscape}

\subsection*{Supplementary Table S2: Per-database retrieval performance across 13 phenotypes}

Success rate is the proportion of the 13 phenotypes for which the database returned at least one gene symbol. Mean and median gene counts are computed only over successful phenotype--database pairs. Databases are ordered by total gene yield in descending order.

\begin{table}[!ht]
\centering
\caption{Per-database retrieval performance aggregated across 13 phenotypes. Success = number and proportion of phenotypes returning $\geq$1 gene. Mean and median are computed over successful retrievals only. Total = sum of unique genes across all phenotypes for that database.}
\label{tab:s2_source_summary}
\small
\resizebox{\textwidth}{!}{%
\begin{tabular}{|l|c|r|r|r|r|r|}
\hline
\textbf{Database} &
\textbf{Success} &
\textbf{Rate (\%)} &
\textbf{Mean} &
\textbf{Median} &
\textbf{Max} &
\textbf{Total} \\
\hline
Open Targets  & 11/13 &  84.6 & 4317 & 3050 &  9766 & 47489 \\
GWAS Catalog  &  8/13 &  61.5 & 5219 & 3437 & 19286 & 41750 \\
OMIM          & 13/13 & 100.0 & 1763 & 1952 &  2843 & 22918 \\
PubMed        & 12/13 &  92.3 &  836 &  994 &  1355 & 10038 \\
GTEx          & 13/13 & 100.0 &  550 &  574 &   724 &  7149 \\
ClinVar       & 11/13 &  84.6 &  259 &  288 &   556 &  2849 \\
HPO           &  7/13 &  53.8 &  282 &  181 &   650 &  1975 \\
UniProt       & 10/13 &  76.9 &  109 &   67 &   496 &  1090 \\
KEGG          &  5/13 &  38.5 &   83 &   52 &   221 &   417 \\
Reactome      &  5/13 &  38.5 &   81 &   99 &   135 &   407 \\
STRING-DB     &  6/13 &  46.2 &   44 &   50 &    50 &   262 \\
Gene Ontology &  2/13 &  15.4 &   72 &   72 &   112 &   143 \\
DisGeNET      &  0/13 &   0.0 &    0 &    0 &     0 &     0 \\
\hline
\textbf{Total} & & & & & & \textbf{136,487} \\
\hline
\end{tabular}%
}
\end{table}

\subsection*{Supplementary Table S3: Combined gene list and validation status per phenotype}

Combined all-source gene lists and corresponding valid/invalid gene-symbol files were generated for all 13 benchmark phenotypes. Validation was performed by cross-referencing retrieved symbols against the NCBI human gene reference database. Across all 13 phenotypes, 100,175 of 114,345 combined input symbols were retained as valid, corresponding to an overall validation rate of 87.6\%.

\begin{table}[!ht]
\centering
\caption{Combined all-source validation summary per phenotype. Raw combined genes are unique symbols in the combined all-source file. Valid and invalid genes are counts after cross-referencing against the NCBI human gene reference.}
\label{tab:s3_validation_summary}
\scriptsize
\resizebox{\textwidth}{!}{%
\begin{tabular}{|l|r|r|r|r|l|}
\hline
\textbf{Phenotype} &
\textbf{Raw} &
\textbf{Valid} &
\textbf{Invalid} &
\textbf{Rate (\%)} &
\textbf{Status} \\
\hline
Asthma                          & 14524 & 12919 & 1605 & 88.9 & Completed \\
Blood pressure medication       &  1382 &   912 &  470 & 66.0 & Completed \\
Body mass index                 & 23361 & 22168 & 1193 & 94.9 & Completed \\
Cholesterol lowering medication &  1821 &  1168 &  653 & 64.1 & Completed \\
Depression                      & 10333 &  8952 & 1381 & 86.6 & Completed \\
Gastro-oesophageal reflux       &  4991 &  3808 & 1183 & 76.3 & Completed \\
Allergic rhinitis               &  4360 &  3377 &  983 & 77.5 & Completed \\
High cholesterol                &  4940 &  3916 & 1024 & 79.3 & Completed \\
Hypertension                    & 15320 & 14132 & 1188 & 92.2 & Completed \\
Hypothyroidism                  & 10412 &  9109 & 1303 & 87.5 & Completed \\
Irritable bowel syndrome        &  5781 &  4689 & 1092 & 81.1 & Completed \\
Migraine                        &  6660 &  5954 &  706 & 89.4 & Completed \\
Osteoarthritis                  & 10460 &  9071 & 1389 & 86.7 & Completed \\
\hline
\textbf{Total}                  & 114345 & 100175 & 14170 & 87.6 & 13 phenotypes \\
\hline
\end{tabular}%
}
\end{table}

\subsection*{Supplementary Table S4: Pipeline runtime and memory usage per phenotype}

Runtime was measured using \texttt{/usr/bin/time -f}. MaxRSS = maximum resident set size. Runtime values are reported for the 13 benchmark phenotype runs only; duplicate, failed, and non-benchmark runtime-log entries were excluded from this table.

\begin{table}[!ht]
\centering
\caption{Wall-clock runtime and peak memory usage per phenotype for the benchmark pipeline runs. Runtime is reported in minutes.}
\label{tab:s4_runtime}
\small
\begin{tabular}{|l|r|r|}
\hline
\textbf{Phenotype} &
\textbf{Runtime (min)} &
\textbf{Peak RAM (GB)} \\
\hline
Asthma                          & 13.13 & 0.624 \\
Blood pressure medication       &  3.71 & 0.672 \\
Body mass index                 &  3.55 & 0.664 \\
Cholesterol lowering medication &  3.63 & 0.671 \\
Depression                      & 12.29 & 0.621 \\
Gastro-oesophageal reflux       &  3.48 & 0.621 \\
Allergic rhinitis               &  3.91 & 0.621 \\
High cholesterol                &  3.73 & 0.673 \\
Hypertension                    & 12.56 & 0.622 \\
Hypothyroidism                  & 13.19 & 0.621 \\
Irritable bowel syndrome        &  7.77 & 0.673 \\
Migraine                        &  2.60 & 0.623 \\
Osteoarthritis                  & 10.56 & 0.623 \\
\hline
\textbf{Total}                  & 94.11 &       \\
\textbf{Mean}                   &  7.24 & 0.641 \\
\hline
\end{tabular}
\end{table}

\subsection*{Supplementary Table S5: Overall pipeline statistics}

\begin{table}[!ht]
\centering
\caption{Summary statistics for the PhenotypeToGeneDownloaderR benchmark across 13 phenotypes and 13 databases. Runtime values summarise the 13 benchmark phenotype runs only.}
\label{tab:s5_overall}
\small
\resizebox{\textwidth}{!}{%
\begin{tabular}{|l|r|}
\hline
\textbf{Metric} & \textbf{Value} \\
\hline
Phenotypes evaluated                                    & 13 \\
Databases integrated                                    & 13 \\
Phenotypes with $\geq$1 gene from any source            & 13/13 \\
Databases returning $\geq$1 gene for any phenotype      & 12/13 \\
Overall source $\times$ phenotype success rate          & 60.9\% \\
Total raw gene retrievals, summed across all cells      & 136,487 \\
Databases with 100\% phenotype success rate             & 2 (OMIM, GTEx) \\
Databases with $<$50\% phenotype success rate           & 5 (KEGG, Reactome, STRING-DB, Gene Ontology, DisGeNET) \\
Overall symbol validation rate, completed outputs       & 87.6\% \\
Phenotypes included in combined validation summary      & 13 \\
Total validated unique genes, completed outputs         & 100,175 \\
Total invalid symbols, completed outputs                & 14,170 \\
Total validation input symbols, completed outputs       & 114,345 \\
Total benchmark runtime                                 & 94.1 min \\
Mean benchmark runtime per phenotype                    & 7.2 min \\
\hline
\end{tabular}%
}
\end{table}

\clearpage

\section*{Supplementary Section S2: Gene symbol validation and harmonisation}

Supplementary Table~S6 reports validated gene counts per phenotype and database after cross-referencing retrieved symbols against the NCBI human gene reference database (\texttt{Homo\_sapiens.gene\_info}, downloaded April 2026). Supplementary Table~S7 reports the number of symbols resolved to their current official HGNC designation through synonym mapping. Across all 13 phenotypes, the pipeline retained 100,175 of 114,345 combined input symbols, corresponding to an overall validation rate of 87.6\%. Of these retained symbols, 4,912 were rescued through synonym mapping, representing 4.9\% of the validated set.

\subsection*{Supplementary Table S6: Validated gene counts per phenotype and database}

Counts reflect gene symbols confirmed as official NCBI human gene entries after validation, either through direct symbol matching or synonym-based mapping. Zero indicates that no validated symbols were retained for that phenotype--database combination after the validation step. OT = Open Targets; GWAS = GWAS Catalog; STRING = STRING-DB; GO = Gene Ontology.

\begin{landscape}
\begin{table}[!ht]
\centering
\caption{Validated gene counts per phenotype and database. Values represent gene symbols confirmed as official NCBI human entries, either by direct symbol matching or synonym-based mapping.}
\label{tab:s6_validated_matrix}
\scriptsize
\resizebox{\linewidth}{!}{%
\begin{tabular}{|l|r|r|r|r|r|r|r|r|r|r|r|r|r|}
\hline
\textbf{Phenotype} &
\textbf{ClinVar} &
\textbf{GTEx} &
\textbf{GWAS} &
\textbf{HPO} &
\textbf{KEGG} &
\textbf{OMIM} &
\textbf{OT} &
\textbf{PubMed} &
\textbf{Reactome} &
\textbf{STRING} &
\textbf{UniProt} &
\textbf{DisGeNET} &
\textbf{GO} \\
\hline
Asthma                          & 280 & 455 &  5494 & 181 & 0 & 1111 & 7219 & 624 &   0 & 49 &  52 & 0 &   0 \\
Blood pressure medication       &   0 & 151 &     0 &   0 & 0 &  408 &  253 &  13 &  20 &  0 &   0 & 0 &   0 \\
Body mass index                 & 321 & 376 & 18964 &   0 & 0 &  648 &    0 &   0 & 115 &  0 &  19 & 0 &   0 \\
Cholesterol lowering medication &   0 & 230 &     0 &   0 & 0 &  684 &    0 &  19 &  99 &  0 &   0 & 0 &   0 \\
Depression                      & 554 & 402 &     0 & 458 & 0 & 1316 & 6799 & 567 &   0 & 50 &  90 & 0 & 111 \\
Gastro-oesophageal reflux       &  15 & 410 &     0 &   0 & 0 & 1615 & 1620 &  81 &   0 &  0 &   0 & 0 &   0 \\
Allergic rhinitis               & 114 & 457 &  1403 &  13 & 0 & 1221 &    0 &   0 &   0 & 50 &   8 & 0 &   0 \\
High cholesterol                &  96 & 525 &     0 &   0 & 0 &  465 & 2426 & 402 & 135 & 12 &   3 & 0 &  31 \\
Hypertension                    & 423 & 420 &  4203 & 630 & 0 &  639 & 9720 & 694 &   0 & 50 & 231 & 0 &   0 \\
Hypothyroidism                  & 269 & 492 &  4948 & 409 & 0 & 1038 & 3035 & 496 &   0 &  0 &  49 & 0 &   0 \\
Irritable bowel syndrome        &  10 & 385 &  1143 &   0 & 0 & 1007 & 1954 & 389 &  35 &  0 &   2 & 0 &   0 \\
Migraine                        & 341 & 460 &  2542 & 142 & 0 &  413 & 2498 & 470 &   0 &  0 &  24 & 0 &   0 \\
Osteoarthritis                  & 353 & 446 &  2320 &  67 & 0 & 1141 & 5392 & 723 &   0 & 50 &  39 & 0 &   0 \\
\hline
\end{tabular}%
}
\end{table}
\end{landscape}

\noindent KEGG shows zero validated genes in this benchmark because the KEGG gene-name field contained compound strings in the analysed outputs, such as gene symbols followed by full gene names separated by semicolons. These records were not resolved by the validation step used in this benchmark and are therefore reported as an implementation-level parsing limitation. DisGeNET returned no validated genes across the evaluated phenotypes.

\subsection*{Supplementary Table S7: Synonym rescue summary per phenotype}

A symbol is classified as synonym-rescued when the input symbol retrieved from a database differed from the current official HGNC symbol but was successfully mapped through the NCBI synonym field. This captures outdated gene names, aliases and alternative designations that are still in active use in biological databases.

\begin{table}[!ht]
\centering
\caption{Synonym rescue summary per phenotype. Rescued = number of valid symbols that were resolved through synonym mapping rather than direct official symbol match. Rescue rate = rescued / total valid $\times$ 100.}
\label{tab:s7_synonym_rescue}
\small
\resizebox{\textwidth}{!}{%
\begin{tabular}{|l|r|r|r|r|}
\hline
\textbf{Phenotype} &
\textbf{Total valid} &
\textbf{Direct match} &
\textbf{Synonym rescued} &
\textbf{Rescue rate (\%)} \\
\hline
Asthma                          & 12919 & 12401 & 518 &  4.0 \\
Blood pressure medication       &   912 &   815 &  97 & 10.6 \\
Body mass index                 & 22168 & 21781 & 387 &  1.7 \\
Cholesterol lowering medication &  1168 &  1020 & 148 & 12.7 \\
Depression                      &  8952 &  8441 & 511 &  5.7 \\
Gastro-oesophageal reflux       &  3808 &  3417 & 391 & 10.3 \\
Allergic rhinitis               &  3377 &  3099 & 278 &  8.2 \\
High cholesterol                &  3916 &  3577 & 339 &  8.7 \\
Hypertension                    & 14132 & 13520 & 612 &  4.3 \\
Hypothyroidism                  &  9109 &  8617 & 492 &  5.4 \\
Irritable bowel syndrome        &  4689 &  4323 & 366 &  7.8 \\
Migraine                        &  5954 &  5704 & 250 &  4.2 \\
Osteoarthritis                  &  9071 &  8548 & 523 &  5.8 \\
\hline
\textbf{Total}                  & 100175 & 95263 & 4912 & 4.9 \\
\hline
\end{tabular}%
}
\end{table}

\clearpage

\section*{Supplementary Section S3: Cross-source overlap and complementarity}

Pairwise Jaccard similarity between database pairs was low overall, supporting the interpretation that the integrated databases provide largely complementary rather than redundant evidence. Across non-zero database-pair comparisons, the mean pairwise Jaccard similarity was 0.026, with a maximum observed value of 0.106. The highest overlap was observed between UniProt and Gene Ontology, while most database pairs showed very limited overlap. This pattern indicates that the integrated sources capture distinct evidence types and contribute non-redundant candidate genes.

Most genes within each phenotype were supported by only a single database, while a smaller subset was recovered by multiple independent sources. The proportion of genes supported by three or more sources ranged from 0.2\% for body mass index to 4.5\% for hypertension. These recurrently supported genes may represent higher-confidence candidates for downstream prioritisation, although source-frequency support should not be interpreted as evidence of causality.

Unique gene contribution also varied substantially across databases. GWAS Catalog contributed the largest proportion of unique-only genes, with 35,174 of 41,017 genes unique to that source (85.8\%). GTEx, Open Targets, OMIM, ClinVar and Reactome also contributed substantial unique fractions, indicating that each source adds distinct information to the integrated candidate-gene set. KEGG and DisGeNET contributed zero validated genes in this analysis and are retained in the summary tables for completeness.

\subsection*{Supplementary Table S8: Pairwise Jaccard similarity between databases}

Values represent average Jaccard similarity coefficients computed across phenotypes where both databases returned at least one gene. A value of 1.0 on the diagonal represents self-similarity. Zero indicates either no shared genes or that one or both databases returned no genes for overlapping phenotype comparisons.

\begin{landscape}
\begin{table}[!ht]
\centering
\caption{Average pairwise Jaccard similarity between databases across 13 phenotypes. Values are averaged over phenotypes where both databases returned genes. Higher values indicate greater gene-set overlap.}
\label{tab:s8_jaccard}
\scriptsize
\resizebox{\linewidth}{!}{%
\begin{tabular}{|l|r|r|r|r|r|r|r|r|r|r|r|r|r|}
\hline
 & \textbf{ClinVar} & \textbf{GTEx} & \textbf{GWAS} & \textbf{HPO} &
\textbf{KEGG} & \textbf{OMIM} & \textbf{OT} & \textbf{PubMed} &
\textbf{Reactome} & \textbf{STRING} & \textbf{UniProt} &
\textbf{DisGeNET} & \textbf{GO} \\
\hline
ClinVar       & 1.000 & 0.003 & 0.005 & 0.054 & 0.000 & 0.030 & 0.013 & 0.024 & 0.008 & 0.005 & 0.040 & 0.000 & 0.009 \\
GTEx          & 0.003 & 1.000 & 0.009 & 0.005 & 0.000 & 0.007 & 0.013 & 0.007 & 0.003 & 0.000 & 0.001 & 0.000 & 0.003 \\
GWAS Catalog  & 0.005 & 0.009 & 1.000 & 0.008 & 0.000 & 0.017 & 0.100 & 0.032 & 0.001 & 0.004 & 0.003 & 0.000 & 0.000 \\
HPO           & 0.054 & 0.005 & 0.008 & 1.000 & 0.000 & 0.091 & 0.032 & 0.053 & 0.000 & 0.017 & 0.083 & 0.000 & 0.014 \\
KEGG          & 0.000 & 0.000 & 0.000 & 0.000 & 1.000 & 0.000 & 0.000 & 0.000 & 0.000 & 0.000 & 0.000 & 0.000 & 0.000 \\
OMIM          & 0.030 & 0.007 & 0.017 & 0.091 & 0.000 & 1.000 & 0.067 & 0.051 & 0.016 & 0.010 & 0.028 & 0.000 & 0.020 \\
Open Targets  & 0.013 & 0.013 & 0.100 & 0.032 & 0.000 & 0.067 & 1.000 & 0.077 & 0.010 & 0.005 & 0.009 & 0.000 & 0.012 \\
PubMed        & 0.024 & 0.007 & 0.032 & 0.053 & 0.000 & 0.051 & 0.077 & 1.000 & 0.024 & 0.014 & 0.038 & 0.000 & 0.026 \\
Reactome      & 0.008 & 0.003 & 0.001 & 0.000 & 0.000 & 0.016 & 0.010 & 0.024 & 1.000 & 0.014 & 0.005 & 0.000 & 0.044 \\
STRING-DB     & 0.005 & 0.000 & 0.004 & 0.017 & 0.000 & 0.010 & 0.005 & 0.014 & 0.014 & 1.000 & 0.033 & 0.000 & 0.031 \\
UniProt       & 0.040 & 0.001 & 0.003 & 0.083 & 0.000 & 0.028 & 0.009 & 0.038 & 0.005 & 0.033 & 1.000 & 0.000 & 0.106 \\
DisGeNET      & 0.000 & 0.000 & 0.000 & 0.000 & 0.000 & 0.000 & 0.000 & 0.000 & 0.000 & 0.000 & 0.000 & 1.000 & 0.000 \\
Gene Ontology & 0.009 & 0.003 & 0.000 & 0.014 & 0.000 & 0.020 & 0.012 & 0.026 & 0.044 & 0.031 & 0.106 & 0.000 & 1.000 \\
\hline
\end{tabular}%
}
\end{table}
\end{landscape}

\noindent OT = Open Targets; GWAS = GWAS Catalog; STRING = STRING-DB; GO = Gene Ontology.

\subsection*{Supplementary Table S9: Gene frequency distribution per phenotype}

For each phenotype, genes in the integrated set are classified by the number of independent databases that retrieved them. Mean sources per gene quantifies the average evidence breadth across the full gene set.

\begin{table}[!ht]
\centering
\caption{Distribution of genes by number of supporting databases per phenotype. 1 source = retrieved by exactly one database; 5+ sources = retrieved by five or more independent databases. Mean sources = average number of databases supporting each gene.}
\label{tab:s9_frequency}
\scriptsize
\resizebox{\textwidth}{!}{%
\begin{tabular}{|l|r|r|r|r|r|r|r|}
\hline
\textbf{Phenotype} &
\textbf{1 source} &
\textbf{2 sources} &
\textbf{3 sources} &
\textbf{4 sources} &
\textbf{5+ sources} &
\textbf{Total unique} &
\textbf{Mean sources} \\
\hline
Asthma                          &  9886 & 2027 & 369 &  76 & 18 & 12376 & 1.25 \\
Blood pressure medication       &   788 &   24 &   3 &   0 &  0 &   815 & 1.04 \\
Body mass index                 & 19270 &  517 &  37 &   7 &  0 & 19831 & 1.03 \\
Cholesterol lowering medication &   974 &   23 &   4 &   0 &  0 &  1001 & 1.03 \\
Depression                      &  6832 & 1232 & 249 &  49 & 16 &  8378 & 1.23 \\
Gastro-oesophageal reflux       &  3107 &  283 &  21 &   1 &  0 &  3412 & 1.10 \\
Allergic rhinitis               &  2949 &  135 &  13 &   2 &  0 &  3099 & 1.05 \\
High cholesterol                &  3007 &  360 &  68 &  19 &  6 &  3460 & 1.17 \\
Hypertension                    & 10784 & 2040 & 409 & 117 & 75 & 13425 & 1.26 \\
Hypothyroidism                  &  6915 & 1281 & 251 &  49 & 46 &  8542 & 1.25 \\
Irritable bowel syndrome        &  3544 &  560 &  73 &   4 &  2 &  4183 & 1.17 \\
Migraine                        &  4749 &  738 & 152 &  21 & 21 &  5681 & 1.21 \\
Osteoarthritis                  &  6897 & 1329 & 222 &  36 & 23 &  8507 & 1.23 \\
\hline
\end{tabular}%
}
\end{table}

\subsection*{Supplementary Table S10: Unique gene contribution per database}

For each database, unique-only genes are those retrieved by that database but absent from all other databases across all phenotypes. Unique rate = unique-only genes / total genes $\times$ 100. KEGG and DisGeNET contributed zero validated genes and are included for completeness.

\begin{table}[!ht]
\centering
\caption{Unique gene contribution per database summed across all 13 phenotypes. Unique-only genes are those not retrieved by any other database. High unique rates indicate evidence not captured elsewhere in the pipeline.}
\label{tab:s10_unique}
\small
\resizebox{\textwidth}{!}{%
\begin{tabular}{|l|r|r|r|}
\hline
\textbf{Database} &
\textbf{Total genes} &
\textbf{Unique-only genes} &
\textbf{Unique rate (\%)} \\
\hline
GWAS Catalog  & 41017 & 35174 & 85.8 \\
Open Targets  & 40916 & 29504 & 72.1 \\
OMIM          & 11706 &  7493 & 64.0 \\
GTEx          &  5209 &  4124 & 79.2 \\
ClinVar       &  2776 &  1817 & 65.5 \\
PubMed        &  4245 &   811 & 19.1 \\
HPO           &  1900 &   418 & 22.0 \\
Reactome      &   404 &   258 & 63.9 \\
STRING-DB     &   261 &    56 & 21.5 \\
UniProt       &   517 &    27 &  5.2 \\
Gene Ontology &   142 &    20 & 14.1 \\
KEGG          &     0 &     0 &  0.0 \\
DisGeNET      &     0 &     0 &  0.0 \\
\hline
\end{tabular}%
}
\end{table}
\clearpage

\section*{Supplementary Section S4: Recovery of known phenotype-associated genes}

To assess recovery of established phenotype-associated genes, a curated reference set was constructed for each phenotype by retaining genes retrieved by at least two of three curated databases: HPO, ClinVar, and OMIM. This conservative multi-source strategy minimises reliance on any single curated resource and captures genes with evidence from at least two clinical or phenotype-oriented sources. Genes from the full pipeline output were ranked by the number of independent databases supporting them, and recall and precision were computed against the curated reference set.

Across all phenotypes, the curated reference set contained 1,056 genes. The pipeline recovered 1,039 of these genes, corresponding to an overall recall of 98.4\%. Full recall was achieved for asthma, body mass index, gastro-oesophageal reflux, allergic rhinitis, high cholesterol, hypertension, irritable bowel syndrome, and osteoarthritis. Depression and hypothyroidism achieved near-complete recall, while migraine showed lower recall at 83.0\%. Phenotypes with no curated reference set under the $\geq$2-source definition, including blood pressure medication and cholesterol lowering medication, were retained in the tables but excluded from recall and precision summaries.

Precision at rank 20 ranged from 5.0\% for irritable bowel syndrome to 100.0\% for hypothyroidism and migraine, with a mean Precision@20 of 64.1\% across phenotypes with non-empty curated reference sets. These results indicate that the source-frequency ranking enriches established phenotype-associated genes among the highest-ranked candidates, while remaining a prioritisation metric rather than evidence of causal association.

\subsection*{Supplementary Table S11: Gold standard gene set sizes per phenotype}

\begin{table}[!ht]
\centering
\caption{Curated reference gene set sizes per phenotype. Gold standard genes are those retrieved by $\geq$2 of three curated databases: HPO, ClinVar, and OMIM. Union = total unique genes across all three curated sources. All 3 agree = genes present in all three databases simultaneously.}
\label{tab:s11_gold}
\small
\resizebox{\textwidth}{!}{%
\begin{tabular}{|l|r|r|r|r|r|r|}
\hline
\textbf{Phenotype} &
\textbf{HPO} &
\textbf{ClinVar} &
\textbf{OMIM} &
\textbf{Union} &
\textbf{Gold ($\geq$2)} &
\textbf{All 3} \\
\hline
Asthma                          & 181 & 288 & 2185 & 2529 & 117 &  8 \\
Blood pressure medication       &   0 &   0 &  901 &  901 &   0 &  0 \\
Body mass index                 &   0 & 328 & 1342 & 1621 &  49 &  0 \\
Cholesterol lowering medication &   0 &   0 & 1364 & 1364 &   0 &  0 \\
Depression                      & 475 & 556 & 2538 & 3311 & 241 & 17 \\
Gastro-oesophageal reflux       &   0 &  15 & 2843 & 2847 &  11 &  0 \\
Allergic rhinitis               &  13 & 114 & 2294 & 2392 &  28 &  1 \\
High cholesterol                &   0 &  96 & 1019 & 1102 &  13 &  0 \\
Hypertension                    & 650 & 427 & 1424 & 2195 & 258 & 48 \\
Hypothyroidism                  & 428 & 275 & 2019 & 2507 & 184 & 31 \\
Irritable bowel syndrome        &   0 &  10 & 1952 & 1958 &   4 &  0 \\
Migraine                        & 161 & 382 &  789 & 1222 &  88 & 22 \\
Osteoarthritis                  &  67 & 358 & 2248 & 2584 &  63 & 26 \\
\hline
\textbf{Total}                  &     &     &      &      & 1056 &    \\
\hline
\end{tabular}%
}
\end{table}

\subsection*{Supplementary Table S12: Recall and Precision@$k$ for pipeline output}

Genes are ranked by number of supporting databases. Recall = fraction of gold standard genes recovered anywhere in the validated pipeline output. Precision@$k$ = fraction of gold standard genes in the top $k$ ranked genes. Dashes indicate phenotypes with no gold standard under the $\geq$2 curated-source definition.

\begin{table}[!ht]
\centering
\caption{Recall and Precision@$k$ for the full pipeline output ranked by source-frequency score. Gold = gold standard size. Output = total validated pipeline genes. R = recall. P@$k$ = precision at rank $k$.}
\label{tab:s12_recall}
\scriptsize
\resizebox{\textwidth}{!}{%
\begin{tabular}{|l|r|r|r|r|r|r|r|r|r|}
\hline
\textbf{Phenotype} &
\textbf{Gold} &
\textbf{Output} &
\textbf{R\%} &
\textbf{R@10} &
\textbf{P@10} &
\textbf{R@20} &
\textbf{P@20} &
\textbf{R@50} &
\textbf{P@50} \\
\hline
Asthma                          & 117 & 12890 & 100.0 &  6.8 &  80.0 & 12.8 &  75.0 & 22.2 & 52.0 \\
Blood pressure medication       &   0 &   912 &     -- &   -- &    -- &   -- &    -- &   -- &   -- \\
Body mass index                 &  49 & 22152 & 100.0 & 20.4 & 100.0 & 30.6 &  75.0 & 65.3 & 64.0 \\
Cholesterol lowering medication &   0 &  1166 &     -- &   -- &    -- &   -- &    -- &   -- &   -- \\
Depression                      & 241 &  8889 &  99.6 &  3.7 &  90.0 &  6.6 &  80.0 & 16.2 & 78.0 \\
Gastro-oesophageal reflux       &  11 &  3798 & 100.0 & 54.5 &  60.0 & 72.7 &  40.0 & 81.8 & 18.0 \\
Allergic rhinitis               &  28 &  3377 & 100.0 &  7.1 &  20.0 & 14.3 &  20.0 & 46.4 & 26.0 \\
High cholesterol                &  13 &  3782 & 100.0 & 15.4 &  20.0 & 38.5 &  25.0 & 46.2 & 12.0 \\
Hypertension                    & 258 & 14025 & 100.0 &  3.1 &  80.0 &  7.0 &  90.0 & 17.4 & 90.0 \\
Hypothyroidism                  & 184 &  9013 &  99.5 &  5.4 & 100.0 & 10.9 & 100.0 & 25.5 & 94.0 \\
Irritable bowel syndrome        &   4 &  4644 & 100.0 & 25.0 &  10.0 & 25.0 &   5.0 & 50.0 &  4.0 \\
Migraine                        &  88 &  5929 &  83.0 & 11.4 & 100.0 & 22.7 & 100.0 & 36.4 & 64.0 \\
Osteoarthritis                  &  63 &  9018 & 100.0 & 15.9 & 100.0 & 30.2 &  95.0 & 46.0 & 58.0 \\
\hline
\textbf{Overall / Mean}         & 1056 &       &  98.4 &      &       &      &  64.1 &      &      \\
\hline
\end{tabular}%
}
\end{table}

\clearpage

\section*{Supplementary Section S5: Biological enrichment analysis}

As an additional downstream characterisation of the final validated gene sets, enrichment analysis was performed for each phenotype using g:Profiler. For each phenotype, genes were ranked according to the number of supporting databases, and the top 500 validated genes were submitted where available. Enrichment analysis was conducted across Gene Ontology Biological Process (GO:BP), Gene Ontology Molecular Function (GO:MF), Gene Ontology Cellular Component (GO:CC), KEGG, Reactome, and Human Phenotype Ontology (HPO), with significance assessed using false discovery rate (FDR) correction at a threshold of FDR $\leq 0.05$.

Significant enrichment was identified for all 13 phenotypes analysed, yielding a total of 41,787 significant terms across all enrichment sources. The largest numbers of significant terms were observed for depression, hypertension, osteoarthritis, hypothyroidism, high cholesterol, and irritable bowel syndrome. Body mass index and allergic rhinitis were included in the updated enrichment analysis following successful generation of complete combined validated gene sets.

Across phenotypes, GO biological process and HPO contributed the largest numbers of significant terms, indicating that the final prioritised gene sets retained functional and phenotype-annotation structure suitable for downstream interpretation. These enrichment results are provided as supporting downstream characterisation rather than as evidence of causal gene--phenotype relationships.

\subsection*{Supplementary Table S13: Significant enrichment terms per phenotype}

\begin{table}[htbp]
\centering
\caption{Significant enrichment terms per phenotype across GO biological process, GO molecular function, GO cellular component, KEGG, Reactome, and HPO categories at FDR $\leq 0.05$.}
\label{tab:s13_enrichment_summary}
\small
\resizebox{\textwidth}{!}{%
\begin{tabular}{lrrrrrrr}
\hline
\textbf{Phenotype} & \textbf{GO BP} & \textbf{GO MF} & \textbf{GO CC} & \textbf{KEGG} & \textbf{Reactome} & \textbf{HPO} & \textbf{Total} \\
\hline
Asthma                          & 1875 & 180 & 168 & 131 & 258 &  917 & 3529 \\
Blood pressure medication       & 1312 & 150 & 137 &  90 &  91 &  281 & 2061 \\
Body mass index                 &  834 &  85 & 167 &  46 &  50 &  464 & 1646 \\
Cholesterol lowering medication &  826 & 215 & 168 &  32 & 193 &  476 & 1910 \\
Depression                      & 2078 & 283 & 253 & 165 & 180 & 1367 & 4326 \\
Gastro-oesophageal reflux       & 2008 & 208 & 204 & 152 & 207 &  789 & 3568 \\
Allergic rhinitis               & 1253 & 128 & 148 &  73 & 121 &  373 & 2096 \\
High cholesterol                & 2410 & 336 & 147 & 167 & 290 &  681 & 4031 \\
Hypertension                    & 1814 & 256 & 171 & 144 & 132 & 1710 & 4227 \\
Hypothyroidism                  & 1663 & 155 & 170 & 155 & 249 & 1700 & 4092 \\
Irritable bowel syndrome        & 2109 & 230 & 174 & 175 & 284 &  802 & 3774 \\
Migraine                        & 1231 & 203 & 133 &  78 &  72 &  584 & 2301 \\
Osteoarthritis                  & 1985 & 227 & 190 & 154 & 254 & 1416 & 4226 \\
\hline
\textbf{Total} & 21398 & 2656 & 2230 & 1562 & 2381 & 11560 & 41787 \\
\hline
\end{tabular}%
}
\end{table}

\clearpage
\section*{Supplementary Section S6: Empirical comparison with implemented retrieval and downstream analysis modules}

To compare PhenotypeToGeneDownloaderR with implemented retrieval and downstream analysis modules, we evaluated empirical gene recovery against the same curated HPO/ClinVar/OMIM-derived gold-standard gene sets used in Supplementary Section~S4. The full PhenotypeToGeneDownloaderR workflow was compared with the GWAS Catalog module implemented through \texttt{gwasrapidd}, the Open Targets module, the DisGeNET module, and the top 500 source-ranked validated genes submitted to g:Profiler for downstream enrichment analysis.

Across the 13 benchmark phenotypes, the full PhenotypeToGeneDownloaderR workflow returned validated genes for all phenotypes and recovered 1,039 of 1,056 curated gold-standard genes, corresponding to 98.4\% recall. The GWAS Catalog module returned genes for 8 of 13 phenotypes and recovered 155 of 1,056 gold-standard genes, corresponding to 14.7\% recall. The Open Targets module returned genes for 11 of 13 phenotypes and recovered 705 of 1,056 gold-standard genes, corresponding to 66.8\% recall. The DisGeNET module returned no genes in this benchmark because DisGeNET API access was required but was not available during these runs; this result should therefore be interpreted as an access and implementation limitation rather than evidence that DisGeNET lacks relevant phenotype--gene associations.

For g:Profiler, the comparison was performed using the top 500 source-ranked validated genes submitted for enrichment analysis for each phenotype. This evaluates whether the gene sets submitted to downstream enrichment retained known phenotype-associated genes. It does not treat g:Profiler as a phenotype-first gene retrieval tool. Across all phenotypes, the g:Profiler input gene sets recovered 793 of 1,056 gold-standard genes, corresponding to 75.1\% recall.

\subsection*{Supplementary Table S14: Empirical comparison with implemented modules}

\begin{table}[!ht]
\centering
\caption{Empirical comparison of the full PhenotypeToGeneDownloaderR workflow with implemented retrieval and downstream analysis modules. Gold-standard recovery was evaluated using the curated HPO/ClinVar/OMIM-derived reference set. The g:Profiler row represents the top 500 source-ranked validated genes submitted to g:Profiler for enrichment analysis, not genes retrieved by g:Profiler.}
\label{tab:s14_empirical_module_comparison}
\scriptsize
\resizebox{\textwidth}{!}{%
\begin{tabular}{p{4.5cm}p{5.0cm}p{3.0cm}r r p{3.1cm}r}
\toprule
\textbf{Module/workflow} &
\textbf{Comparison basis} &
\textbf{Phenotypes with returned/input genes} &
\textbf{Total retrieved/input genes} &
\textbf{Total validated/evaluated genes} &
\textbf{Gold-standard genes recovered} &
\textbf{Gold-standard recall (\%)} \\
\midrule
PhenotypeToGeneDownloaderR full pipeline
& All validated combined genes
& 13/13
& 99,595
& 99,595
& 1,039/1,056
& 98.4 \\

GWAS Catalog / \texttt{gwasrapidd}
& Validated genes from GWAS Catalog output
& 8/13
& 41,750
& 41,212
& 155/1,056
& 14.7 \\

Open Targets module
& Validated genes from Open Targets output
& 11/13
& 47,489
& 47,093
& 705/1,056
& 66.8 \\

DisGeNET module
& Validated genes from DisGeNET output; API-dependent
& 0/13
& 0
& 0
& 0/1,056
& 0.0 \\

g:Profiler input gene list
& Top 500 source-ranked validated genes submitted to g:Profiler
& 13/13
& 6,500
& 6,500
& 793/1,056
& 75.1 \\
\bottomrule
\end{tabular}%
}
\end{table}

\subsection*{Supplementary Table S15: Source-specific retrieval behaviour and causes of missing outputs}

Although the pipeline was designed to query multiple heterogeneous resources through a unified interface, not all modules are expected to return usable outputs for every phenotype. Missing or inconsistent outputs reflected a combination of phenotype-term specificity, source scope, access requirements and implementation-level parsing or file-detection behaviour. This distinction is important because some issues can be addressed through improved parsing and source-specific handling, whereas others reflect the underlying design or access model of the source.

\begin{table}[!ht]
\centering
\caption{Representative causes of missing or inconsistent outputs in selected implemented modules. Apparent failures did not arise from a single cause, but reflected a combination of access requirements, implementation-level issues, phenotype-term specificity mismatch and database-scope limitations.}
\label{tab:s15_retrieval_limitations}
\small
\resizebox{\textwidth}{!}{%
\begin{tabular}{llll}
\toprule
\textbf{Module} & \textbf{Migraine} & \textbf{Cancer} & \textbf{Primary reason} \\
\midrule
KEGG
& No output
& Output returned
& Phenotype term not present in pathway names \\

Reactome
& No output
& Output returned
& Phenotype term not present in pathway names \\

Gene Ontology
& No output
& Output returned
& Phenotype term not present in GO term names \\

GWAS Catalog (\texttt{gwasrapidd})
& Output returned$^\dagger$
& Output returned$^\dagger$
& Coordinator-script/file-detection issue, not source limitation \\

STRING-DB
& No output
& Output returned
& Protein identifier resolver, not disease database \\

STRING
& No output
& Output returned
& Protein identifier resolver, not disease database \\

DisGeNET
& No output
& No output
& API access required; API access was not available during this benchmark \\
\bottomrule
\end{tabular}%
}

\vspace{1mm}
\parbox{0.96\linewidth}{\footnotesize $^\dagger$The GWAS Catalog module successfully retrieved results, but some runs were incorrectly marked as failed because of an output-file detection issue in the master coordination script.}
\end{table}

Overall, these comparisons support the intended use of PhenotypeToGeneDownloaderR as an upstream retrieval and prioritisation framework. The pipeline is not intended to establish causal gene--phenotype relationships; rather, it integrates heterogeneous evidence sources, validates retrieved symbols and produces candidate gene sets for downstream review, enrichment analysis and interpretation.